\documentclass[12pt]{article}
\usepackage{epsfig}
\usepackage{times}
\def\beq{\begin{equation}}
\def\eeq{\end{equation}}
\def\bea{\begin{eqnarray}}
\def\eea{\end{eqnarray}}
\def\d{\displaystyle}

\def\nn{\nonumber}
\def\Eq#1{Eq.~(\ref{#1})}

\begin{document}
\begin{titlepage}

\renewcommand{\thefootnote}{\fnsymbol{footnote}}
\begin{flushright}
    IFIC/08-49  \\ arXiv:0809.3354 [hep-ph]
\end{flushright}
\par \vspace{10mm}
  
\begin{center}

{\Large \bf Massive color-octet bosons and the charge asymmetries 
of top quarks at hadron colliders}

\vspace{8mm}

{\bf Paola Ferrario~\footnote{E-mail: paola.ferrario@ific.uv.es}} and
{\bf Germ\'an Rodrigo~\footnote{E-mail: german.rodrigo@ific.uv.es}}

\vspace{5mm}
Instituto de F\'{\i}sica Corpuscular, 
CSIC-Universitat de Val\`encia, \\
Apartado de Correos 22085, 
E-46071 Valencia, Spain. \\

\vspace{5mm}
\end{center}

\par \vspace{2mm}
\begin{center} {\large \bf Abstract} \end{center}
\begin{quote}
Several models predict the existence of heavy colored resonances 
decaying to top quarks in the TeV energy range that might be 
discovered at the LHC. In some of those 
models, moreover, a sizable charge asymmetry of top versus 
antitop quarks might be generated. The detection of these 
exotic resonances, however, requires selecting data samples 
where the top and the antitop quarks are highly boosted, which is 
experimentally very challenging. We asses that the measurement 
of the top quark charge asymmetry at the LHC is very sensitive 
to the existence of excited states of the gluon with axial-vector 
couplings to quarks. We use a toy model with general flavour 
independent couplings, and show that a signal can be detected 
with relatively not too energetic top and antitop quarks.
We also compare the results with the asymmetry predicted by 
QCD, and show that its highest statistical significance 
is achieved with data samples of top-antitop quark pairs 
of low invariant masses. 
\end{quote}

\vspace*{\fill}
\begin{flushleft}
     IFIC/08-49 \\
     September 19, 2008
\end{flushleft}
\end{titlepage}

\setcounter{footnote}{1}
\renewcommand{\thefootnote}{\fnsymbol{footnote}}


\section{Introduction}

The Large Hadron Collider (LHC) will start-up very soon
colliding protons to protons. 
In a first run, at a center of mass energy of $\sqrt{s}=10$~TeV, 
about $20$~pb$^{-1}$ of data are expected to be collected. 
The 2009 run will operate at the full $\sqrt{s}=14$~TeV design energy
with an initial low luminosity of ${\cal L}=10^{33}$cm$^{-2}$s$^{-1}$
(equivalent to $10$~fb$^{-1}$/year integrated luminosity).
The production cross section of top-antitop quark pairs at LHC
is about $430$~pb at $10$~TeV, and $950$~pb 
at 14~TeV~\cite{Cacciari:2008zb}. The LHC will produce
in the first phase of operation a sample of $t\bar t$-pairs 
equivalent to the sample already collected at Tevatron 
during its whole life, and millions of top-antitop quark pairs 
in the next run at $14$~TeV.
This will allow not only to measure better some of the properties 
of the top quark, such as mass and cross section, but also 
to explore with unprecedented huge statistics the existence 
of new physics at the TeV energy scale in the top quark sector.

\begin{figure}[th]
\begin{center}
\includegraphics[width=9cm]{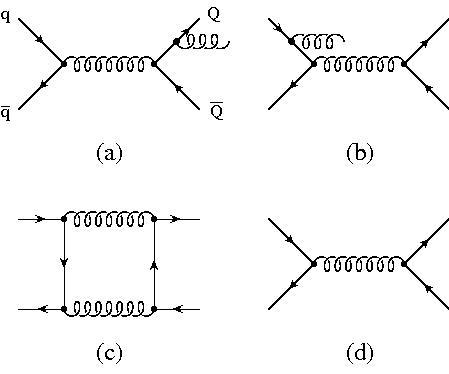}
\caption{Origin of the QCD charge asymmetry in hadroproduction
of heavy quarks: interference of final-state
(a) with initial-state (b) gluon bremsstrahlung,
plus interference of the double virtual gluon exchange (c)
with the Born diagram (d). Only representative diagrams are shown.}
\label{fig:qqbar}
\end{center}
\end{figure}

At leading order in the strong coupling $\alpha_s$ the differential 
distributions of top and antitop quarks are identical. 
This feature changes, however, due to higher order 
corrections~\cite{mynlo}, which predict at ${\cal O}(\alpha_s^3)$
a charge asymmetry of top versus antitop quarks.
A similar effect leads also to a strange-antistrange
quark asymmetry, $s(x)\neq \bar{s}(x)$, through next-to-next-to-leading
(NNLO) evolution of parton densities~\cite{Catani:2004nc}.
At Tevatron, the charge asymmetry is equivalent to a 
forward--backward asymmetry because the top and the antitop 
single inclusive distributions are related by 
$N_{\bar t} (\cos \theta) = N_{t} (-\cos \theta)$
through CP invariance of QCD.  
The inclusive charge asymmetry receives contributions from two reactions:
radiative corrections to quark-antiquark annihilation (Fig.~\ref{fig:qqbar})
and interference between different amplitudes contributing
to gluon-quark scattering $gq \to t \bar{t}q$ and 
$g\bar{q} \to t \bar{t}\bar{q}$. The latter contribution is, 
in general, much smaller than the former. 
Gluon-gluon fusion, which represents only $15\%$ of all the 
events at Tevatron, remains charge symmetric.
QCD predicts that the size of the inclusive charge asymmetry 
is 5 to 8\%~\cite{mynlo,Antunano:2007da,Bowen:2005ap}, 
with top quarks (antitop quarks)
more abundant in the direction of the incoming proton (antiproton).
The prediction for the charge asymmetry is, furthermore, robust 
with respect to the higher-order perturbative corrections generated 
by threshold resummation~\cite{Almeida:2008ug}. The  
forward--backward asymmetry of the exclusive process 
$p\bar{p} \to t\bar{t}+$ jet receives, however, large higher order 
corrections~\cite{ttjetnlo}.

At LHC, the total forward--backward asymmetry vanishes trivially
because the proton-proton initial state is symmetric. Nevertheless,
a charge asymmetry is still visible in suitably defined 
distributions~\cite{mynlo}. In contrast with Tevatron, 
top quark production at LHC is dominated by gluon-gluon 
fusion ($84~\%$ at $10$~TeV, and $90~\%$ at $14$~TeV), which
is charge symmetric under higher order corrections. 
The charge antisymmetric contributions to top quark 
production are thus screened at LHC 
due to the prevalence of gluon-gluon fusion. 
This is the main handicap for that measurement.
The amount of events initiated by gluon-gluon collisions can nevertheless
be suppressed with respect to the $q\bar q$ and $gq(\bar q)$ processes, 
the source of the charge asymmetry, by introducing a lower cut 
on the invariant mass of the top-antitop quark system $m_{t\bar t}$;
this eliminates the region of lower longitudinal momentum 
fraction of the colliding partons, 
where the gluon density is much larger than the quark densities. 
The charge asymmetry of the selected data samples is then enhanced,
although at the price of lowering the statistics. This is, in principle, 
not a problem at LHC, where the high luminosity will compensate 
by far this reduction. 

Several models predict the existence of heavy colored resonances 
decaying to top quarks that might be observed at the 
LHC~\cite{chiralcolor,Bagger:1987fz,FBmodels,Choudhury:2007ux,KK,Randall:1999ee,Dicus:2000hm,Agashe:2006hk,Lillie:2007yh,Lillie:2007ve,Djouadi:2007eg,Agashe:2007jb,Carone:2008rx,Frederix:2007gi}. 
Those resonances will appear as a peak in the invariant mass
distribution of the top-antitop quark pair located at the mass of the new 
resonance. Colored resonances are fairly broad:
$\Gamma_G/m_G = {\cal O}(\alpha_s) \sim 10\%$.
Present lower bounds on their mass are about $1$~TeV.
The latest exclusion limit by CDF~\cite{masslimits} at 95\% C.L. is 
$260$~GeV$ < m_G < 1.250$~TeV for axigluons and flavor-universal colorons
(with $\cot \theta=1$ mixing of the two $SU(3)$).

Some of those exotic gauge bosons, such as the 
axigluons~\cite{chiralcolor,Bagger:1987fz}, 
might generate at tree-level a charge asymmetry too through 
the interference with the $q\bar q \to t \bar t$ 
SM amplitude~\cite{Antunano:2007da,FBmodels,Choudhury:2007ux}.
Gluon-gluon fusion to top quarks stays, at first order, 
unaltered by the presence of new interactions because a pair 
of gluons do not couple to a single extra resonance 
in this kind of models~\cite{Bagger:1987fz,Dicus:2000hm}. 

To discover those resonances, hence, it is necessary to select 
top-antitop quark pair events with large invariant masses; i.e. in 
the vicinity of the mass of the new resonance. 
A sizable charge asymmetry can also be obtained only if 
gluon-gluon fusion is sufficiently suppressed, i.e. 
at large values of $m_{t\bar t}$. 
Because the top quarks of those data samples will be produced
highly boosted, they will be observed as a single monojet. 
The standard reconstruction algorithms that are based on the reconstruction 
of the decay products, however, loose efficiency very rapidly at high 
transverse momentum. For $p_T > 400$~GeV new identification 
techniques are necessary. This has motivated many recent 
investigations~\cite{Kaplan:2008ie,Thaler:2008ju,Vos,Almeida:2008yp} aimed at 
distinguishing top quark jets from the light quark QCD background by
exploiting the jet substructure, without identifying the decay products. 

In this paper we argue that for a measurement of the top quark 
charge asymmetry at LHC it is not necessary to select events 
with very large invariant masses of the top-antitop quark pairs. 
We show that the highest statistical significance occurs with 
moderate selection cuts. Indeed, we find that the measurement 
of the charge asymmetry induced by QCD is better suited in the 
region of low top-antitop quark pair invariant masses. The 
higher statistics in this region compensates the smallness 
of the charge asymmetry. We also investigate the charge 
asymmetry generated by the exchange of a heavy color-octet 
resonance. We study the scenario where the massive extra 
gauge boson have arbitrary flavour independent vector and axial-vector 
couplings to quarks. This includes the case of the axigluon that 
we have already analyzed in Ref.~\cite{Antunano:2007da}.
We first show the constraints that the recent 
measurements at Tevatron of the forward--backward or charge asymmetries 
impose over the parameter space, and then extend the analysis to LHC. 
We show that the selections cuts can be tuned such that the maximum 
statistical significance is obtained with a cut on the 
invariant mass of the top-antitop quark pair at roughly half of 
the mass of the heavy resonance. Due to this fact  
the measurement of the charge asymmetry has potentially 
a better sensitivity to higher masses of the exotic resonances
than the direct measurement of the dijet distribution. 

The outline of the paper is as follows. In Section~\ref{sec:tevatron}
we review the most recent measurements of the top quark charge 
asymmetries at Tevatron, and compare the QCD prediction for the 
asymmetries with the charge asymmetry generated by the exchange of 
a color-octet boson with flavour independent vector and 
axial-vector couplings to quarks. In Section~\ref{sec:QCDLHC}
we evaluate at the LHC the top quark charge asymmetry, as 
predicted by QCD, in a given finite interval of rapidity, and study its 
size and statistical significance as a function of the cut 
in the invariant mass of the top-antitop quark pair. 
In Section~\ref{sec:octetLHC}, we extend to the LHC the analysis 
of the charge asymmetry in the toy model used for Tevatron, and 
show that a detection of those resonances is possible with 
relatively low cuts on the invariant mass of the top-antitop quark pair; 
at values much smaller that the resonance mass.

\section{Top quark charge asymmetry at Tevatron}
\label{sec:tevatron}

The forward--backward asymmetry of top quarks has already been measured
at Tevatron~\cite{newcdf,d0,Weinelt:2006mh,Hirschbuehl:2005bj,Schwarz:2006ud}.
The latest CDF analysis~\cite{newcdf}, based on $1.9$~fb$^{-1}$
integrated luminosity,
provides two different measurements in the lepton+jets channel. 
The first measurement is made in the laboratory frame, and gives 
\beq
A_{\rm FB}^{p \bar p} = \frac{N_t (\cos \theta >0)-N_t (\cos \theta <0)}
{N_t(\cos \theta >0)+N_t(\cos \theta <0)}
= 0.17 \pm 0.07~(\rm{stat}) \pm 0.04~(\rm{sys})~,
\label{eq:newcdf1}
\eeq
where $\theta$ is the angle between the top quark and the proton beam. 
The second measurement exploits the Lorentz invariance of the  
difference between the $t$ and $\bar t$ rapidities, $\Delta y=y_t-y_{\bar t}$, 
which at LO is related to the top quark production angle $\alpha$
in the $t\bar t$ rest frame by~\cite{Hirschbuehl:2005bj}:
\beq
\Delta y = 2 \tanh^{-1} \left( \beta \cos \alpha \right)~, 
\eeq
where $\beta=\sqrt{1-4m_t^2/\hat s}$ is the top quark velocity.
The asymmetry in this variable is
\beq
A_{\rm FB}^{t \bar t} = 
\frac{N_{\rm{ev.}}(\Delta y >0)-N_{\rm{ev.}}(\Delta y <0)}
{N_{\rm{ev.}}(\Delta y >0)+N_{\rm{ev.}}(\Delta y  <0)}
= 0.24 \pm 0.13~(\rm{stat}) \pm 0.04~(\rm{sys})~.
\label{eq:newcdf2}
\eeq
The measurement at D0~\cite{d0} with $0.9$~fb$^{-1}$
integrated luminosity gives for the uncorrected asymmetry
\beq
A^{\rm{obs}}_{\rm FB} = 0.12 \pm 0.08~(\rm{stat}) \pm 0.01~(\rm{sys})~.
\eeq
Like CDF, this analysis uses $y_t-y_{\bar t}$ as sensitive variable.
In Ref.~\cite{d0} upper limits on $t\bar t+X$ production via a $Z'$
resonance are also provided.
Measurements of the exclusive asymmetry of the four- and
five-jet samples are also given by both experiments~\cite{newcdf,d0}.

\begin{figure}[ht]
\begin{center}
\includegraphics[width=4.cm]{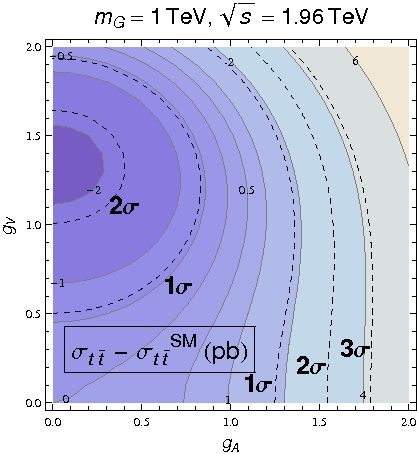} 
\includegraphics[width=4.cm]{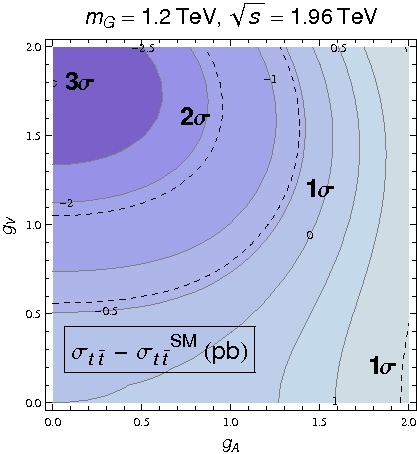} 
\includegraphics[width=4.cm]{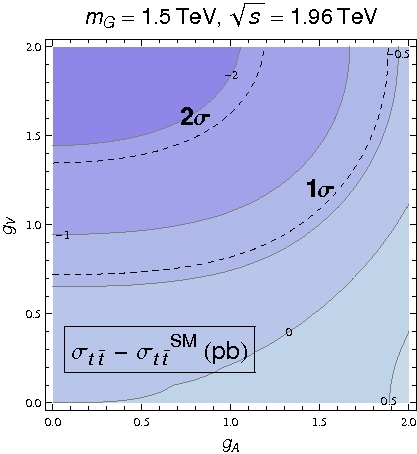} 
\caption{Top quark cross-section at Tevatron in the bidimensional $g_V$-$g_A$ 
plane for different values of the resonance mass.}
\label{fig:tevatron_cross}
\end{center}
\end{figure}
\begin{figure}[ht]
\begin{center}
\includegraphics[width=4.cm]{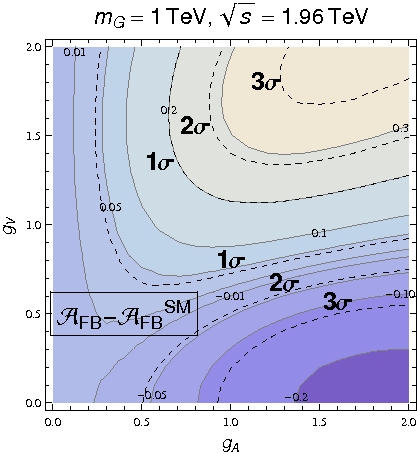} 
\includegraphics[width=4.cm]{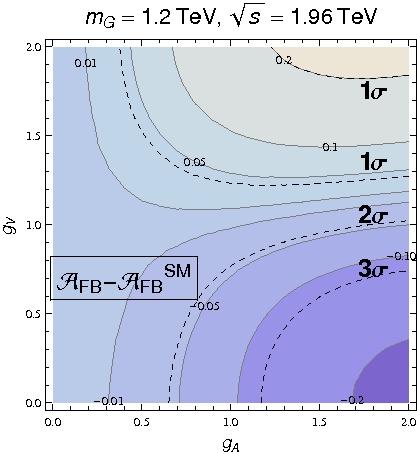} 
\includegraphics[width=4.cm]{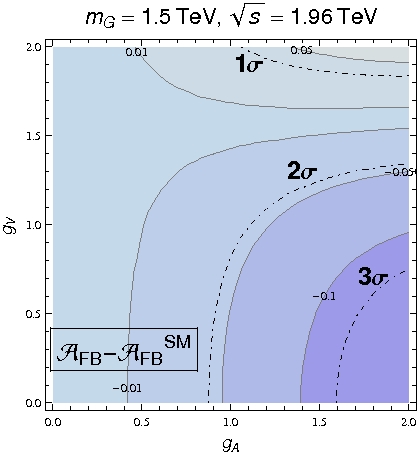} 
\caption{Forward-backward asymmetry at Tevatron in the bidimensional 
$g_V$-$g_A$ plane for different values of the resonance mass.}
\label{fig:tevatron_FB}
\end{center}
\end{figure}
\begin{figure}[ht]
\begin{center}
\includegraphics[width=4.cm]{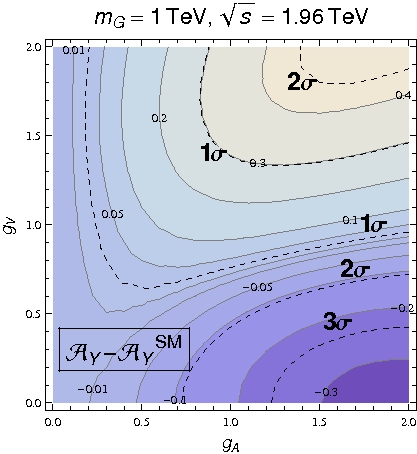} 
\includegraphics[width=4.cm]{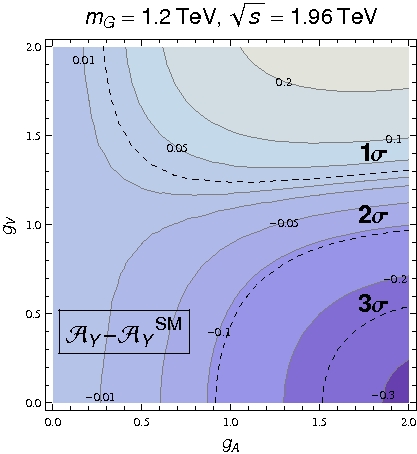} 
\includegraphics[width=4.cm]{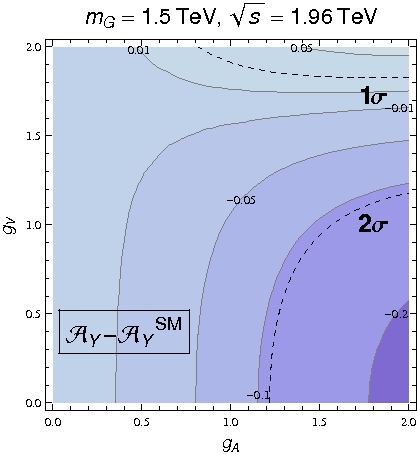} 
\caption{Pair asymmetry at Tevatron in the bidimensional 
$g_V$-$g_A$ plane for different values of the resonance mass.}
\label{fig:tevatron_pair}
\end{center}
\end{figure}

The corresponding theoretical predictions are~\cite{Antunano:2007da}
\beq
{\cal A} =\frac{N_t(y\ge 0)-N_{\bar{t}}(y\ge 0)}
{N_t(y\ge 0)+N_{\bar{t}}(y\ge 0)} = 0.051(6)~,
\label{ourprediction}
\eeq
for the inclusive charge asymmetry, or forward--backward asymmetry 
(${\cal A} = A_{\rm FB}^{p \bar p}$), and 
\beq
{\cal A}_Y = 
\frac{\d \int dY (N_\mathrm{ev.}(y_t>y_{\bar{t}}) - N_\mathrm{ev.}(y_t <y_{\bar{t}}))}
{\d \int dY (N_\mathrm{ev.}(y_t>y_{\bar{t}}) + N_\mathrm{ev.}(y_t<y_{\bar{t}}))}
=  0.078(9)~,
\label{eq:pair}
\eeq
for the integrated pair asymmetry, which is 
defined through the average rapidity $Y=\frac{1}{2}(y_t + y_{\bar t})$.
The differential pair asymmetry is almost flat in the average rapidity, 
and amounts to about $7\%$ for any value of $Y$.
The corresponding integrated asymmetry is 
equivalent to the integrated forward--backward asymmetry 
in the $t\bar t$ rest frame: ${\cal A}_Y=A_{\rm FB}^{t \bar t}$.
The pair asymmetry is larger than the forward--backward 
asymmetry ${\cal A}$ because events 
where both $t$ and $\bar{t}$ are produced with positive and 
negative rapidities in the laboratory frame do not contribute 
to the integrated forward--backward asymmetry, while they do 
contribute to the pair asymmetry. 
The experimental measurements of the top quark asymmetry
in \Eq{eq:newcdf1} and \Eq{eq:newcdf2},
although compatible with the corresponding 
theoretical predictions
in \Eq{ourprediction} and \Eq{eq:pair}, respectively,
are still statistically dominated. 

We shall consider in the following the production 
of heavy color-octet boson resonances decaying to top-antitop quark 
pairs with arbitrary vector and axial-vector couplings to quarks. 
The corresponding differential cross section is given 
in \Eq{eq:bornqq} of Appendix~\ref{ap:born}.
The charge asymmetry is built up from the two contributions
of the differential partonic cross section that are odd in the polar angle.
The first one arises from the interference with the gluon amplitude, 
and is proportional to the product of the axial-vector couplings of the 
light and the top quarks. This contribution, provided that the product 
of couplings is positive, is negative in the forward direction 
for invariant masses of the top-antitop quark pair below the resonance mass, 
and changes sign above.
This leads at Tevatron to a preference for the emission of the top quarks 
in the direction of the incoming light antiquarks (antiprotons) in most of the 
kinematic phase-space, and then to a negative forward--backward asymmetry.  
The second contribution, arising from the squared amplitude of the heavy 
resonance, although always positive for positive couplings,
is suppressed with respect to the contribution of the interference 
term by two powers of the resonance mass.
For large values of the couplings,
however, it might compensate the interference contribution, 
then leading to a positive forward--backward asymmetry,
because it is enhanced by the product of the vector couplings. 
Indeed, for 
\beq
{\hat s} = s' \equiv \frac{m_G^2}{1+2 \, g_V^q \, g_V^t}~,
\label{eq:sprime}
\eeq
the two odd terms cancel to each other, and above that value 
the contribution to the forward--backward asymmetry becomes 
positive.

To simplify our analysis we consider that the vector 
and axial-vector couplings, which are normalized to 
the strong coupling $\alpha_s$, are flavour independent:
$g_V^q = g_V^t = g_V$, and $g_A^q = g_A^t = g_A$, where $q$
labels the coupling of the excited gluon to light quarks, 
and $t$ to top quarks. 
The axigluon of chiral color theories~\cite{chiralcolor,Bagger:1987fz}, 
for example, is given by $g_V=0$ and $g_A=1$.  
We study how the production cross section and the charge asymmetry 
vary depending on the vector and axial-vector couplings, which 
we take in the range $[0,2]$. Within this range the perturbative 
expansion is still reliable. Moreover, in the flavour independent 
scenario the sign of the couplings is not relevant.

\begin{figure}[th]
\begin{center}
\includegraphics[width=7cm,height=5cm]{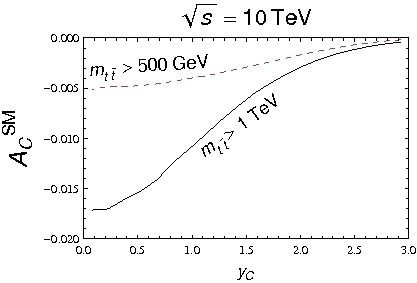} 
\includegraphics[width=7cm,height=5cm]{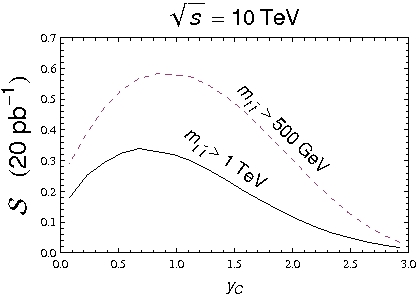}  \\
\includegraphics[width=7cm,height=5cm]{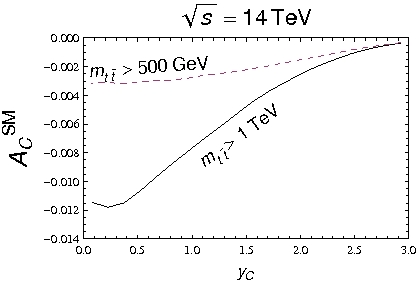} 
\includegraphics[width=7cm,height=5cm]{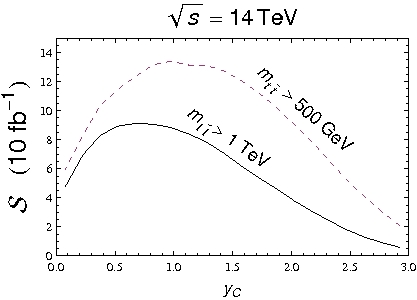}  
\caption{Central charge asymmetry at LHC as predicted by QCD, as a function of 
the maximum rapidity $y_C$ (left plots), and corresponding statistical 
significance (right plots), for two different cuts on the top-antitop 
quark pair invariant mass.}
\label{fig:LHC_QCD_yc}
\end{center}
\end{figure}

Results for the difference between the production cross section 
of the excited gluon and the SM prediction in the $(g_V, g_A)$ 
plane are presented in Fig.~\ref{fig:tevatron_cross} for 
different values of the resonance mass. 
In all our analysis, we use the MRST 2004 parton distribution 
functions~\cite{Martin:2006qz}, 
and we set the renormalization and factorization 
scales to $\mu=m_t$, with 
$m_{t}=170.9 \pm 1.1~(\mathrm{stat}) \pm 
1.5~(\mathrm{sys})$~GeV~\cite{topmass}. 
For comparison, we also overimpose in 
Fig.~\ref{fig:tevatron_cross} the $1$, $2$ and $3$ sigma contours obtained 
from the experimental measurement 
$\sigma_{t\bar t} = 7.3 \pm 0.9$~(pb)~\cite{topcross},
and the SM prediction  
$\sigma_{t\bar t}^{\rm{NLO}} = 6.7 \pm 0.8$~(pb)~\cite{Cacciari:2003fi}.
Similar plots are presented in Fig.~\ref{fig:tevatron_FB}
and Fig.~\ref{fig:tevatron_pair} for the forward--backward 
asymmetry and pair asymmetry, respectively. 
The sigma contours are calculated 
from the experimental measurement in \Eq{eq:newcdf1} and 
from the theoretical prediction in \Eq{ourprediction} for the 
forward--backward asymmetry, and from \Eq{eq:newcdf2}
and \Eq{eq:pair} for the pair asymmetry. 
At $90\%$ C.L. the plots of the production cross section 
and the asymmetries exclude complementary regions 
of the parameter space. While the production cross
section excludes the corner with large vector couplings
$g_V$ and low axial-vector coupling $g_A$, the forward--backward
and the pair asymmetry exclude the corners with higher 
axial-vector couplings and either low or high vector 
couplings. This is not surprising, because the terms 
of the differential cross section in \Eq{eq:bornqq} that are
even in the polar angle contribute exclusively 
to the integrated cross section, 
while the odd terms contribute to the charge asymmetry only, 
and they are proportional to different combinations
of the vector and axial-vector couplings. 
The exclusion regions are, as expected, 
smaller for higher values of the resonance mass.

\section{QCD induced charge asymmetry at LHC}
\label{sec:QCDLHC}

\begin{figure}[th]
\begin{center}
\includegraphics[width=7cm,height=5cm]{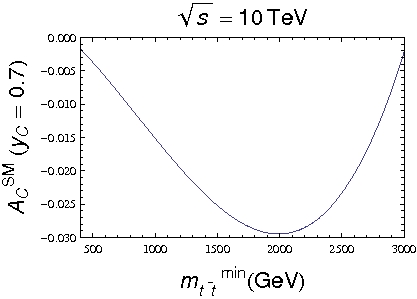} 
\includegraphics[width=7cm,height=5cm]{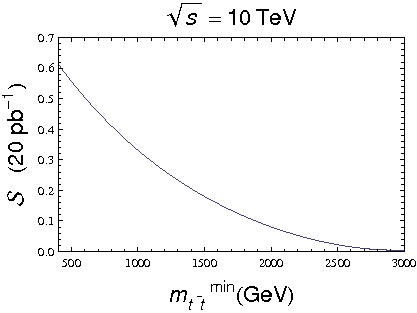} \\
\includegraphics[width=7cm,height=5cm]{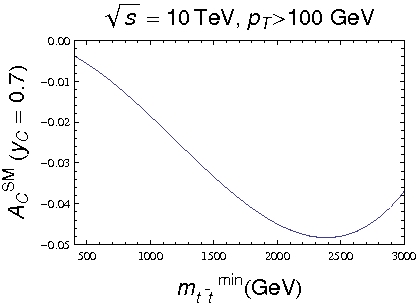} 
\includegraphics[width=7cm,height=5cm]{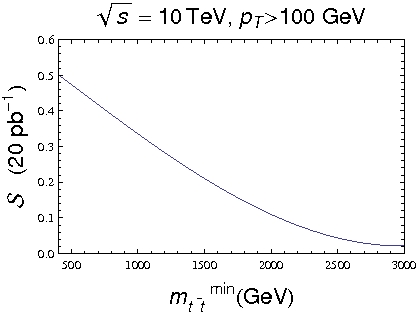} 
\caption{Central charge asymmetry and statistical significance 
at LHC from QCD, as a function of the cut $m_{t\bar t}^{\rm{min.}}$, 
for $10$~TeV energy.}
\label{fig:LHC_QCD_mtt_10TeV}
\end{center}
\end{figure}
\begin{figure}[th]
\begin{center}
\includegraphics[width=7cm,height=5cm]{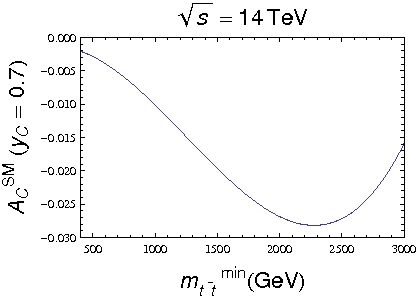} 
\includegraphics[width=7cm,height=5cm]{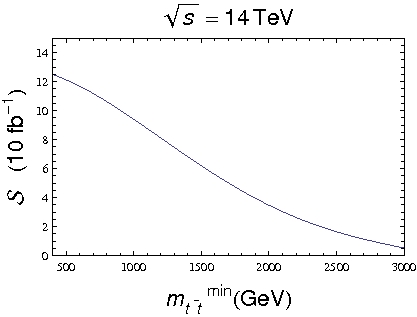} \\
\includegraphics[width=7cm,height=5cm]{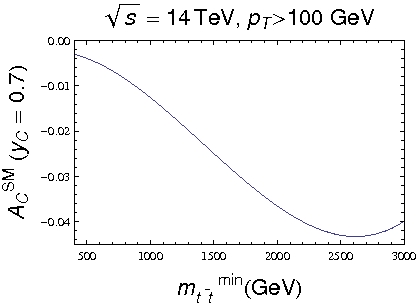} 
\includegraphics[width=7cm,height=5cm]{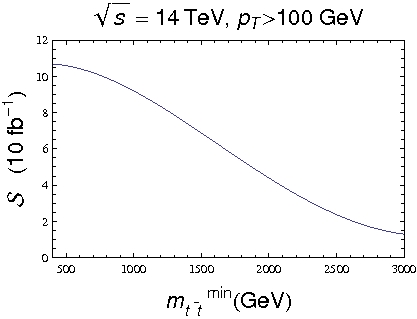} 
\caption{Central charge asymmetry and statistical significance 
at LHC from QCD, as a function of the cut $m_{t\bar t}^{\rm{min.}}$, 
for $14$~TeV energy.}
\label{fig:LHC_QCD_mtt_14TeV}
\end{center}
\end{figure}

Top quark production at LHC is forward--backward symmetric in the 
laboratory frame as a consequence of the symmetric colliding 
proton-proton initial state. The charge asymmetry can be studied 
nevertheless by selecting appropriately chosen kinematic regions.
The production cross section of top quarks is, however,  
dominated by gluon-gluon fusion and thus the charge asymmetry 
generated from the $q\bar{q}$ and $gq$ ($g\bar{q}$) reactions is
small in most of the kinematic phase-space. 

Nonetheless, QCD predicts at LHC a slight preference 
for centrally produced antitop quarks, with top quarks more abundant 
at very large positive and negative rapidities~\cite{mynlo}. 
The difference between the single particle inclusive distributions 
of $t$ and $\bar {t}$ quarks can be understood easily. 
Production of $t\bar{t}(g)$ is dominated by initial quarks with 
large momentum fraction and antiquarks with small momentum 
fraction, while QCD predicts that top (antitop) quarks are preferentially 
emitted into the direction of the incoming quarks (antiquarks)
in the partonic rest frame.
Due to the boost into the laboratory frame the top quarks
are then produced dominantly in the forward and backward directions,
while antiquarks are more abundant in the central region. 

We select events in a given range of rapidity 
and define the integrated charge asymmetry in 
the central region as~\cite{Antunano:2007da}:
\beq
A_C(y_C) = \frac{N_t(|y|\le y_C)-N_{\bar{t}}(|y|\le y_C)}
{N_t(|y|\le y_C)+N_{\bar{t}}(|y|\le y_C)}~.\label{eq:acyc}
\eeq
The central asymmetry $A_C(y_C)$ obviously vanishes if the 
whole rapidity spectrum is integrated, while a non-vanishing 
asymmetry can be obtained over a finite interval of rapidity. 
We also perform a cut on the invariant mass of the top-antitop quark 
pair, $m_{t\bar t} > m_{t\bar t}^{\rm{min}}$,
because that region of the phase space is more sensitive 
to the quark-antiquark induced events rather than the 
gluon-gluon ones, so that the asymmetry is enhanced.
The main virtue of the central asymmetry is that it vanishes 
exactly for parity-conserving processes.

\begin{figure}[th]
\begin{center}
\includegraphics[width=7cm,height=5cm]{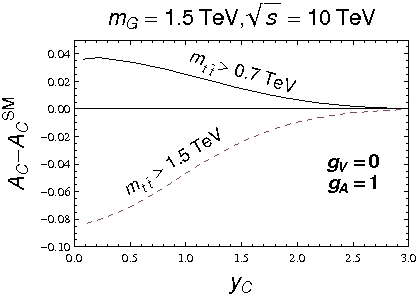}
\includegraphics[width=7cm,height=5cm]{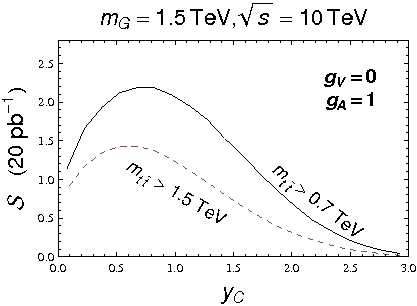} \\
\includegraphics[width=7cm,height=5cm]{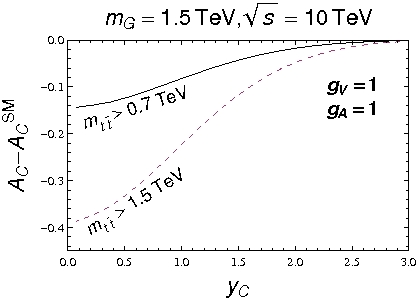}
\includegraphics[width=7cm,height=5cm]{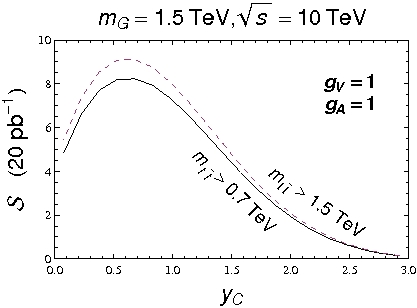}
\caption{Central charge asymmetry (left plots) and statistical 
significance (right plots) at LHC as a function of the maximum 
rapidity, for $10$ TeV energy and two different cuts on the 
top--antitop quark invariant mass. 
Two different sets of couplings are shown.}
\label{fig:LHC_KK_yc_10TeV}
\end{center}
\end{figure}

In Fig.~\ref{fig:LHC_QCD_yc}~(left plots) we show the central charge asymmetry 
at $\sqrt{s}=10$~TeV and $14$~TeV as a function of the maximum rapidity $y_C$ 
for two different values 
of the cut on the invariant mass of the top-antitop quark pair
$m_{t\bar t}> 500~$GeV, and $1$~TeV, respectively. 
As expected, the central charge asymmetry is negative, is 
larger for larger values of the cut $m_{t\bar t}^{\rm{min}}$, and vanishes 
for large values of $y_C$. 
We also show in Fig.~\ref{fig:LHC_QCD_yc}~(right plots) 
the corresponding statistical significance ${\cal S}$ 
of the measurement, defined as 
\beq
{\cal S}^{\rm{SM}} = A_C^{\rm{SM}} \, \sqrt{(\sigma_t+\sigma_{\bar t})^{\rm{SM}} 
\, {\cal L}} = \frac{N_t-N_{\bar t}}{\sqrt{N_t+N_{\bar t}}}~, 
\eeq
where ${\cal L}$ denotes the total integrated luminosity for which we 
take ${\cal L}=20$~pb$^{-1}$ at $\sqrt{s}=10$~TeV 
and ${\cal L}=10$~fb$^{-1}$ at $\sqrt{s}=14$~TeV. 
The maximum significance is reached for both running energies 
at $y_C=1$ for $m_{t\bar t}> 500~$GeV, and 
$y_C=0.7$ for $m_{t\bar t}>1$~TeV. Surprisingly, although the 
size of the asymmetry is greater for the larger value of 
$m_{t\bar t}^{\rm{min}}$, its statistical significance is higher 
for the lower cut. This is a very interesting 
feature because softer top and antitop quarks should be identified 
more easily than the very highly boosted ones.
The statistical significance for the run at $\sqrt{s}=10$~TeV is, 
however, rather small.

We now fix the value of the maximum rapidity to $y_C=0.7$ and study 
the size of the asymmetry and its statistical significance as a 
function of $m_{t\bar t}^{\rm{min}}$.
Our results are shown in Fig.~\ref{fig:LHC_QCD_mtt_10TeV}
for $\sqrt{s}=10$~TeV and ${\cal L}=20$~pb$^{-1}$
and in Fig.~\ref{fig:LHC_QCD_mtt_14TeV}
for $\sqrt{s}=14$~TeV and ${\cal L}=10$~fb$^{-1}$.
We also compare the results with and without introducing 
a cut in the transverse momenta of the heavy quarks:
$p_{T} > 100$~GeV. The latter cut 
produce a significant effect only at very large 
values of the invariant mass $m_{t\bar t}^{\rm{min}}$, 
above $2$~TeV, where the statistical significance 
is, however, small. In all cases, the asymmetry 
increases for larger values of $m_{t\bar t}^{\rm{min}}$, 
while the statistical significance is larger without 
introducing any selection cut. Note that the size of the asymmetry 
decreases again above $m_{t\bar t}^{\rm{min}}=2.5$~TeV
because in that region the $gq(\bar q)$ events compensate 
the asymmetry generated by the $q\bar q$ events; their 
contributions are of opposite sign. 
The statistical significance for the initial run at $\sqrt{s}=10$~TeV is 
again too small. Although we have not taken into account experimental 
efficiencies, we can conclude that $10$~fb$^{-1}$ of data at the design 
energy of the LHC seems to be enough for a clear measurement of the 
QCD asymmetry. 

\begin{figure}[th]
\begin{center}
\includegraphics[width=7cm,height=5cm]{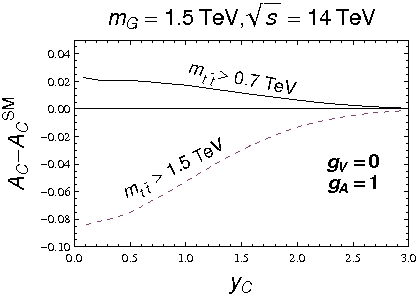}
\includegraphics[width=7cm,height=5cm]{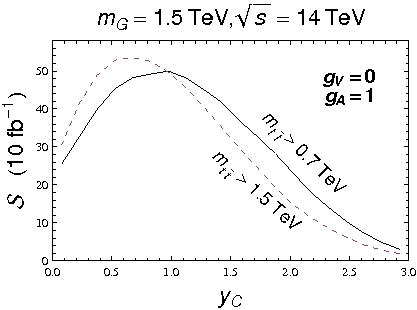} \\
\includegraphics[width=7cm,height=5cm]{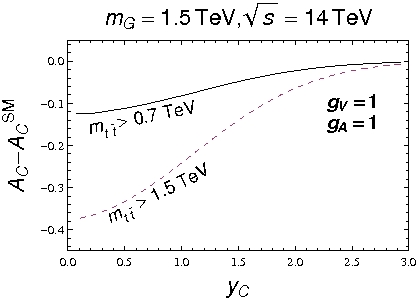} 
\includegraphics[width=7cm,height=5cm]{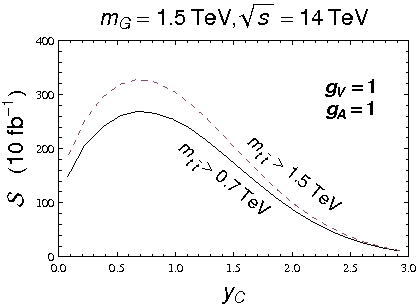}
\caption{Central charge asymmetry (left plots) and statistical 
significance (right plots) at LHC as a function of the maximum 
rapidity, for $10$ TeV energy and two different cuts on the 
top-antitop quark invariant mass. 
Two different sets of couplings are shown.}
\label{fig:LHC_KK_yc_14TeV}
\end{center}
\end{figure}

\section{Charge asymmetry of color-octet resonances at LHC}
\label{sec:octetLHC}

\begin{figure}[th]
\begin{center}
\includegraphics[width=6.cm]{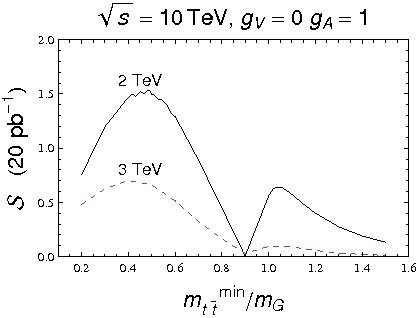} 
\includegraphics[width=6.cm]{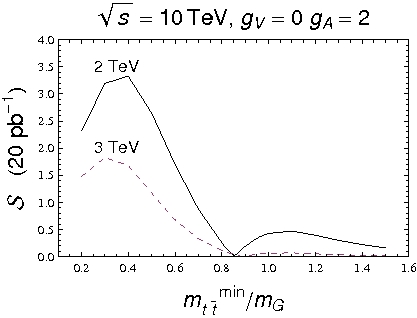}  \\
\includegraphics[width=6.cm]{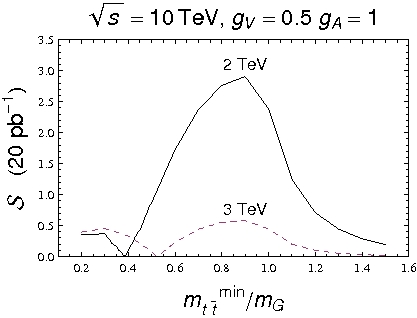} 
\includegraphics[width=6.cm]{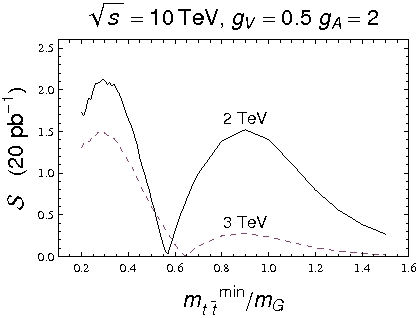} \\
\includegraphics[width=6.cm]{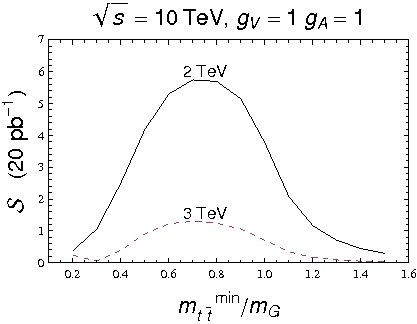} 
\includegraphics[width=6.cm]{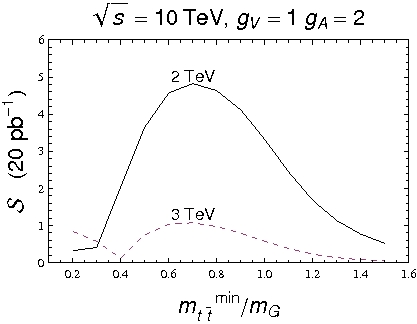}  \\
\includegraphics[width=6.cm]{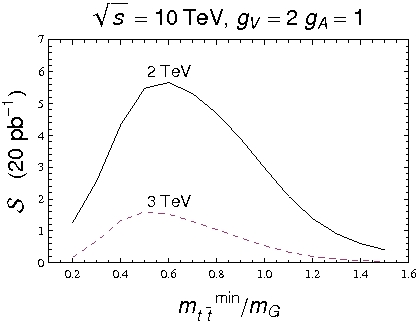} 
\includegraphics[width=6.cm]{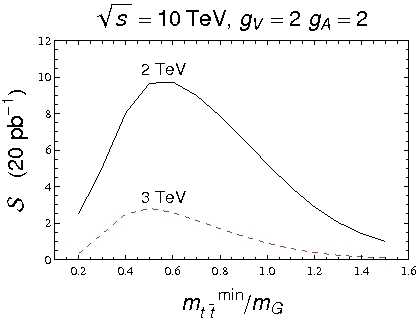} 
\caption{Statistical significance at LHC for $10$ TeV energy and 
different sets of $g_A$, $g_V$ as a function of the cut on the 
top-antitop quark pair invariant mass for $m_G=2$ and $3$~TeV.}
\label{fig:LHC_KK_gvga_10TeV}
\end{center}
\end{figure}

\begin{figure}[th]
\begin{center}
\includegraphics[width=6.cm]{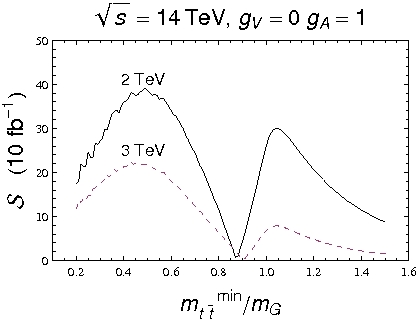} 
\includegraphics[width=6.cm]{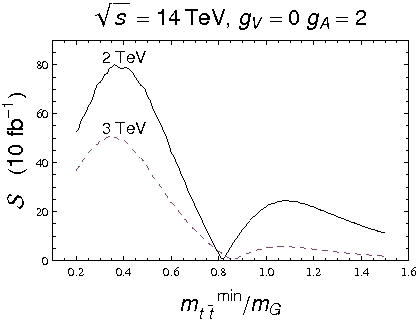}  \\
\includegraphics[width=6.cm]{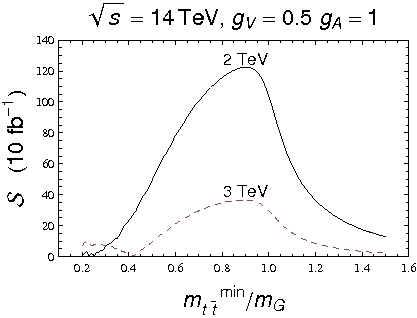} 
\includegraphics[width=6.cm]{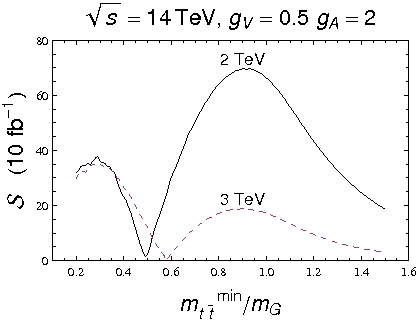} \\
\includegraphics[width=6.cm]{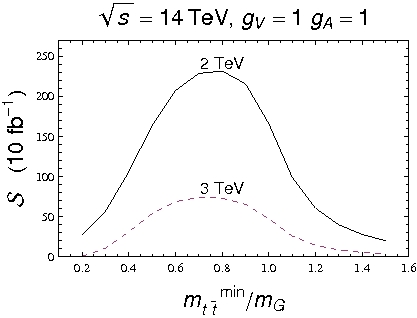} 
\includegraphics[width=6.cm]{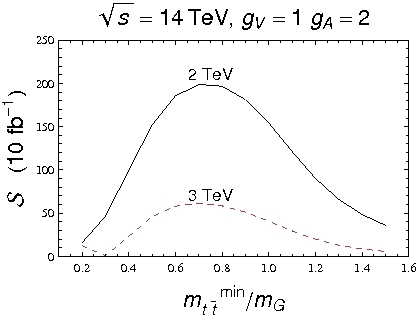}  \\
\includegraphics[width=6.cm]{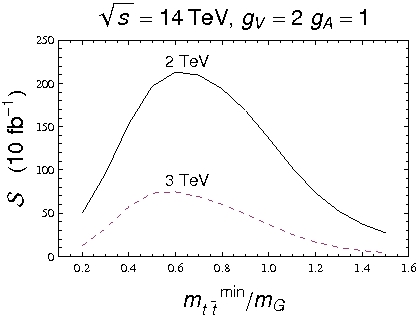} 
\includegraphics[width=6.cm]{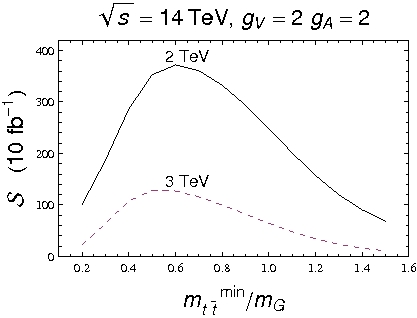} 
\caption{Statistical significance at LHC for $14$ TeV energy and 
different sets of $g_A$, $g_V$ as a function of the cut on the 
top-antitop quark pair invariant mass for $m_G=2$ and $3$~TeV.}
\label{fig:LHC_KK_gvga_14TeV}
\end{center}
\end{figure}

\begin{figure}[th]
\begin{center}
\includegraphics[width=4.cm]{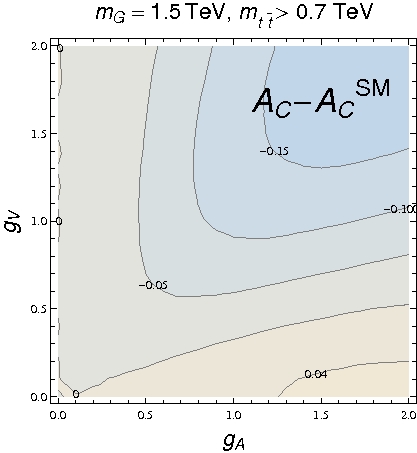} 
\includegraphics[width=4.cm]{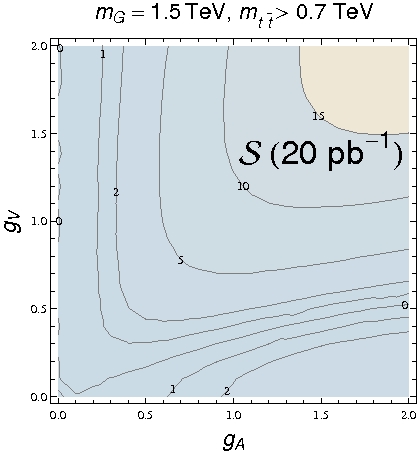} 
\includegraphics[width=4.cm]{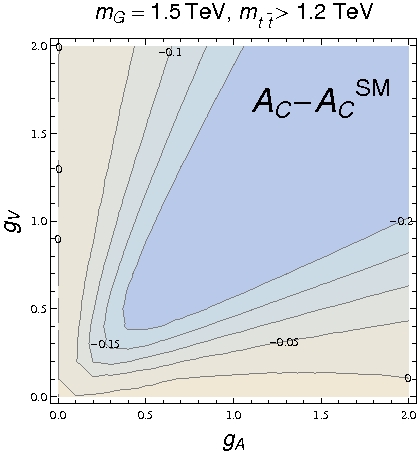} 
\includegraphics[width=4.cm]{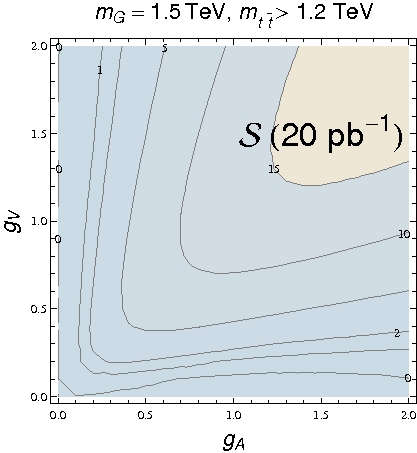} \\
\includegraphics[width=4.cm]{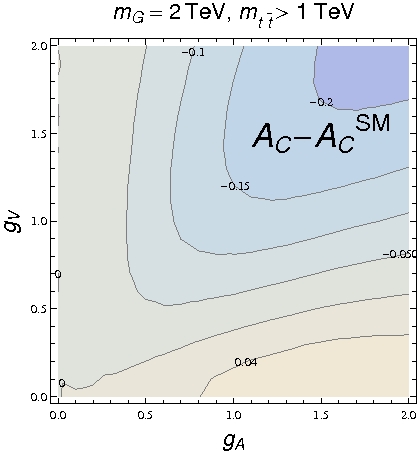} 
\includegraphics[width=4.cm]{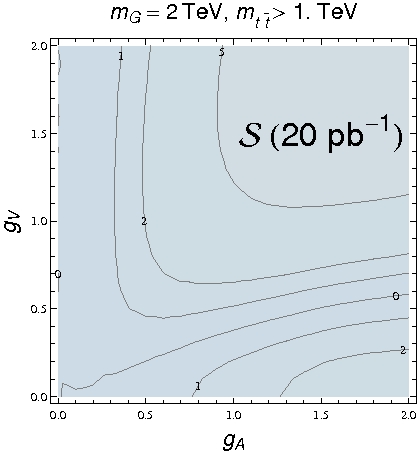} 
\includegraphics[width=4.cm]{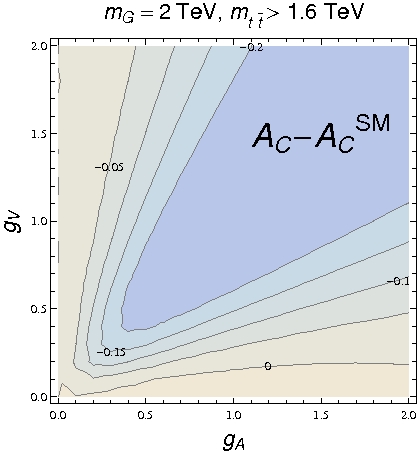} 
\includegraphics[width=4.cm]{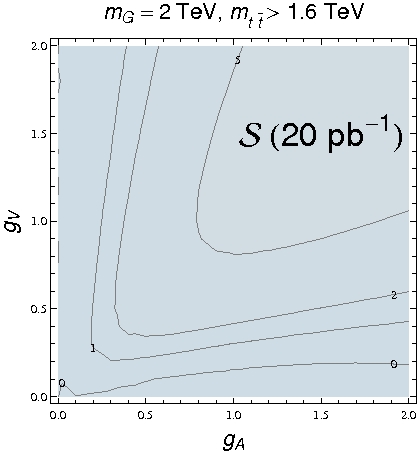} \\
\includegraphics[width=4.cm]{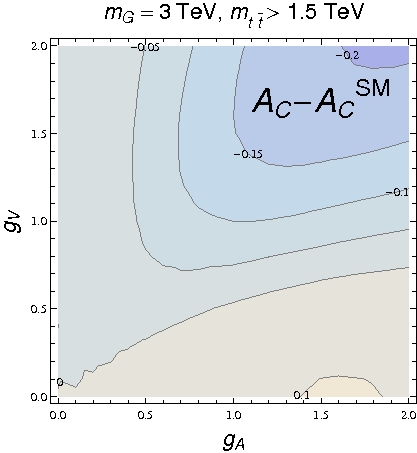} 
\includegraphics[width=4.cm]{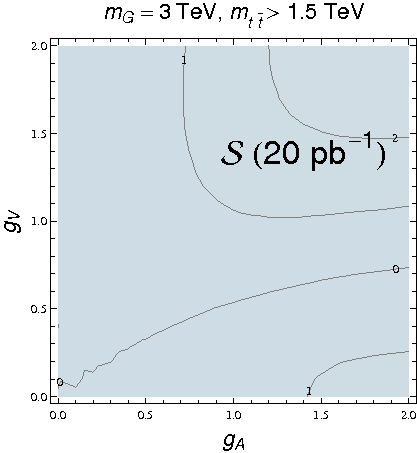} 
\includegraphics[width=4.cm]{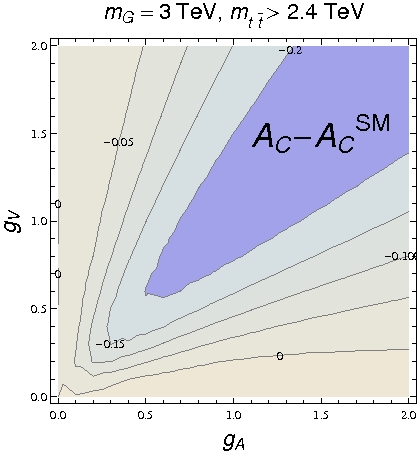} 
\includegraphics[width=4.cm]{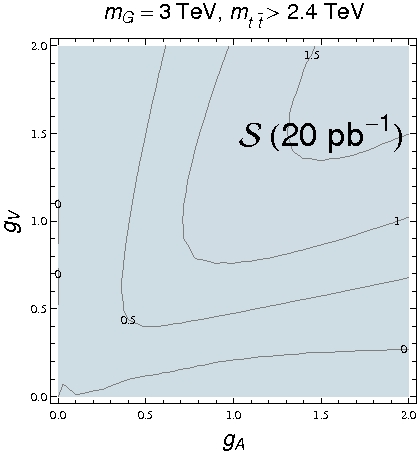} 
\caption{Central charge asymmetry and statistical significance 
at LHC in the $g_A$-$g_V$ plane for $10$ TeV energy, 
for different values of the resonance mass and the cut 
on the top-antitop quark pair invariant mass.}
\label{fig:LHC_KK_2d_10TeV}
\end{center}
\end{figure}

\begin{figure}[th]
\begin{center}
\includegraphics[width=4.cm]{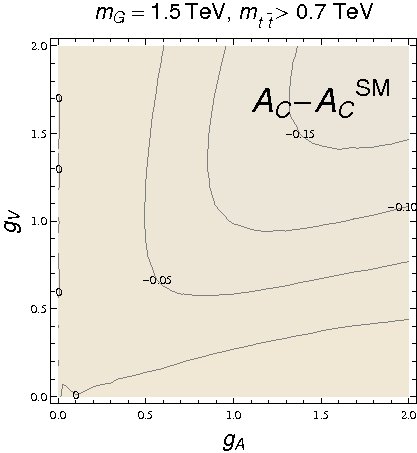} 
\includegraphics[width=4.cm]{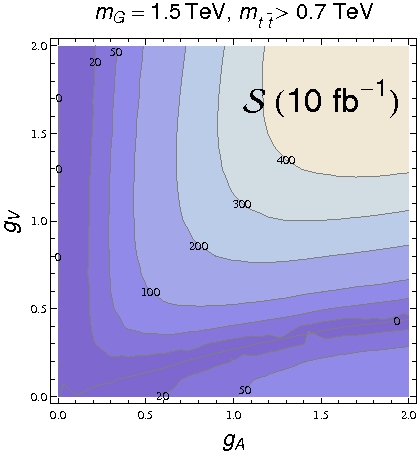} 
\includegraphics[width=4.cm]{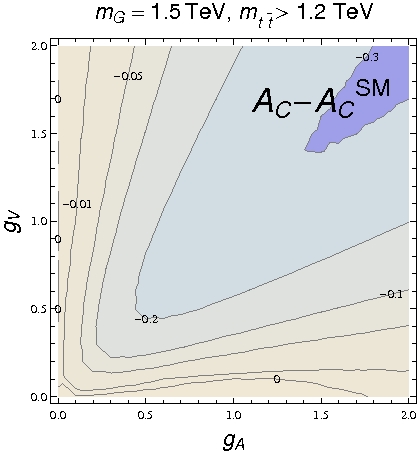} 
\includegraphics[width=4.cm]{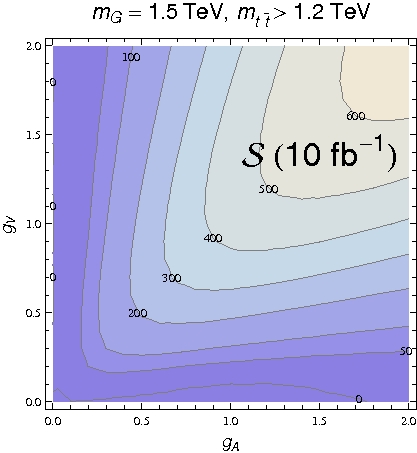} \\
\includegraphics[width=4.cm]{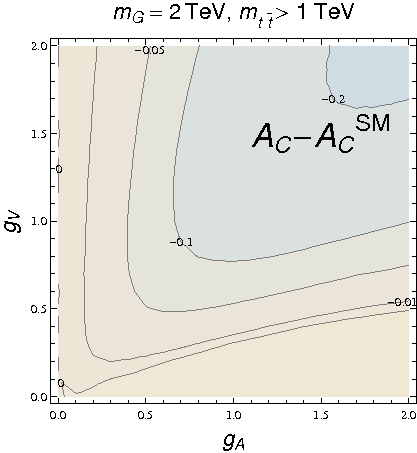} 
\includegraphics[width=4.cm]{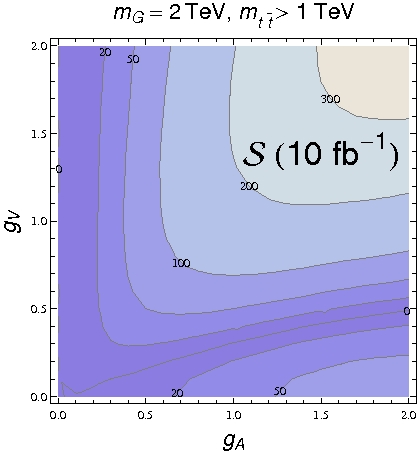} 
\includegraphics[width=4.cm]{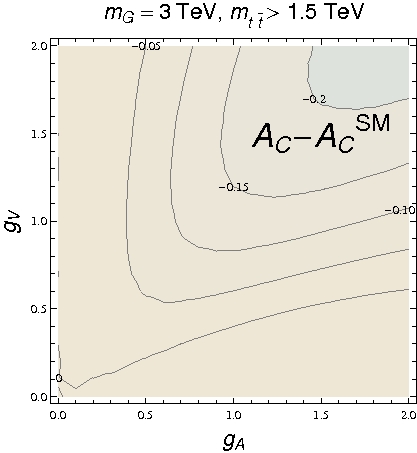} 
\includegraphics[width=4.cm]{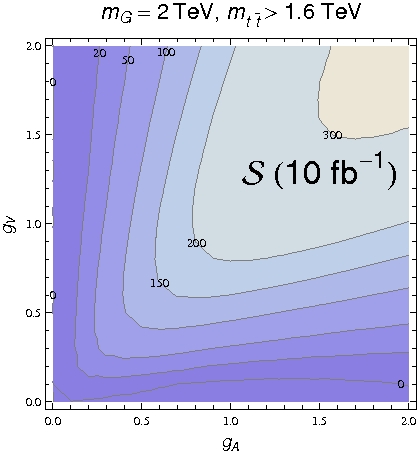} \\
\includegraphics[width=4.cm]{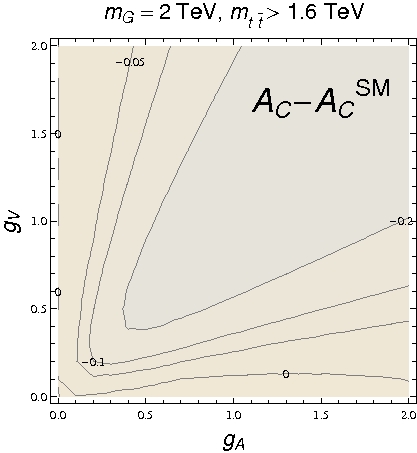} 
\includegraphics[width=4.cm]{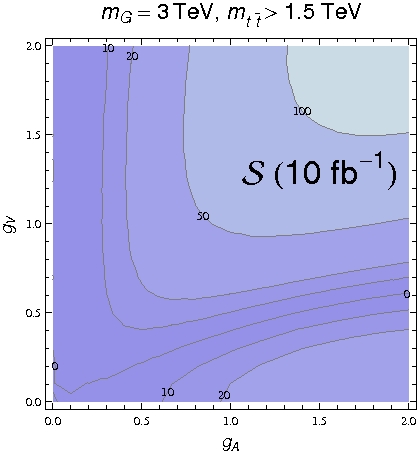} 
\includegraphics[width=4.cm]{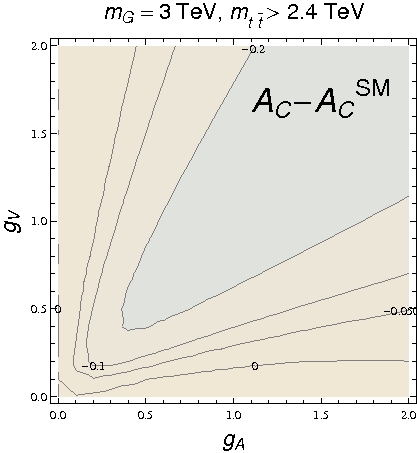} 
\includegraphics[width=4.cm]{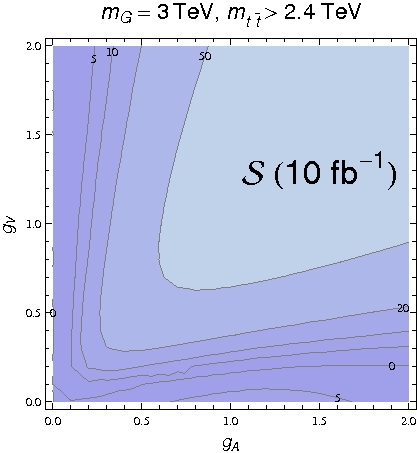} 
\caption{Central charge asymmetry and statistical significance 
at LHC in the $g_A$-$g_V$ plane for $14$ TeV energy, 
for different values of the resonance mass and the cut
on the top-antitop quark pair invariant mass.}
\label{fig:LHC_KK_2d_14TeV}
\end{center}
\end{figure}

Like for Tevatron in Section~\ref{sec:tevatron}, 
we study here the charge asymmetry produced at LHC 
by the decay to top quarks of a color-octet resonance, 
in the scenario where the vector $g_V^{q(t)}$ and 
axial-vector $g_A^{q(t)}$ couplings are flavour independent. 
We evaluate the central asymmetry in \Eq{eq:acyc}, and 
its statistical significance, defined as
\beq
{\cal S}^{\rm G} = 
\frac{A_C^{\rm{G+SM}}-A_C^{\rm{SM}}}{\sqrt{1-(A_C^{\rm{SM}})^2}} 
\, \sqrt{(\sigma_t+\sigma_{\bar t})^{\rm{SM}} \, {\cal L}}
\simeq \frac{(N_t-N_{\bar t})^{\rm{G}}}
{\sqrt{(N_t+N_{\bar t})^{\rm{G+SM}}}}~,
\eeq
for different values of the couplings and the kinematical cuts. 

We should mention that in the most popular models of 
warped extra dimensions
the Kaluza-Klein excitations of the 
gluon couple identically to the left-handed and the right-handed 
light quarks, and these couplings are different only for the 
third generation. A charge asymmetry, or correspondingly 
a central asymmetry, can not be generated 
in this kind of models by the production mechanism. 
An asymmetry will arise, however, in the decay products 
of the top quark. The polarization asymmetry from the angular 
distribution of the positron from the top quark decay 
has been investigated for example in~\cite{Agashe:2006hk}. 
The analysis of the decay products is, however, beyond the scope 
of this paper. The scenario presented here includes, however, the extra 
dimensional model presented in~\cite{Carone:2008rx} as a particular 
subcase. 

When a heavy color-octet boson resonance is produced 
considerations similar to those in Section~\ref{sec:tevatron} lead 
to predict a positive central asymmetry for values of the cut in 
the invariant mass of the top-antitop quark pair below the mass of the 
resonance and a negative asymmetry above. This is true as far as
the interference term has a greater relevance than the squared amplitude 
of the exotic resonance. If this is the case, 
a higher number of antitop quarks will be  
emitted in the direction of the incoming quarks, and
once the boost into the laboratory frame is performed
(cf. discussion in Section~\ref{sec:QCDLHC}), a
higher number of top quark will be found in the central region, 
so that the central asymmetry is positive.
Since for high values of the cut the sign of the
interference term changes, the asymmetry will become negative, 
and then it has to vanish at a certain intermediate 
value of that cut, close and below the resonance mass. 
 
Under these conditions, we expect to find two maxima in the statistical 
significance as a function of $m_{t\bar{t}}^{\rm{min.}}/m_{G}$.
Starting from the threshold, where the asymmetry is small because
the gluon-gluon fusion process dominates there, the size of the 
central asymmetry will grow by increasing $m_{t \bar t}^{\rm{min.}}$,
as the quark-antiquark annihilation process becomes more
and more important. Since the asymmetry induced by the excited gluon 
will vanish at a certain critical point, its statistical significance
will do as well, and will reach a maximum at an intermediate value 
between that critical point and the threshold. 
Above the critical point, the asymmetry becomes negative and 
its statistical significance increases again, until the event yield 
becomes too small. A second maximum in the statistical significance 
will be generated there. 

For certain values of the vector couplings, however, 
the critical partonic invariant mass defined in 
\Eq{eq:sprime} can be located at a rather low 
scale. In this case, the central asymmetry generated by 
the exotic resonance will be negative exclusively, 
and we will find only one maximum in the 
statistical significance.

In our first analysis we shall determine the value of 
the maximum rapidity $y_C$ that maximizes the statistical significance. 
We fix the resonance mass at $1.5$ TeV,
and impose two different cuts on the invariant mass 
of the top-antitop quark pair, namely $m_{t\bar t}> 700 \;\mathrm{GeV}$ 
and $m_{t\bar t}> 1.5\; \mathrm{TeV}$. 
We choose two different combinations of the vector and axial-vector  
couplings $g_V$ and $g_A$.
In Figs. \ref{fig:LHC_KK_yc_10TeV} and \ref{fig:LHC_KK_yc_14TeV}, 
we present the results obtained for the central asymmetry and the 
statistical significance for $g_V=0$, $g_A=1$ and $g_V=g_A=1$ 
for both values of the centre-of-mass energy, $10$ and $14$ TeV,
respectively. 
We notice that for the first choice of the parameters, 
namely $g_V=0$, $g_A=1$, the central asymmetry suffers a change 
of sign by passing from the lower cut to the higher one. 
This means that it will vanish for a given value of the cut, 
thus making the statistical significance vanishing also. 

By looking at the corresponding significance we find that 
$y_C=0.7$ is a good choice in all cases.
Thus, we use this value to find the best cut for the
top-antitop quark pair invariant mass. 
In order to do that, we choose several 
values of the parameters and we study the trend of the 
significance as a function of $m_{t\bar t}^{\rm{min}}/m_{G}$. 
The results are shown in Figs. \ref{fig:LHC_KK_gvga_10TeV} 
and \ref{fig:LHC_KK_gvga_14TeV}. The optimal cuts depend, 
of course, on the values of the vector and axial-vector 
couplings, but either $m_{t\bar{t}}^{\rm{min}}/m_{G}=0.5$ 
or $m_{t\bar{t}}^{\rm{min}}/m_{G}=0.8$ provide a reasonable 
statistical significance for almost all the 
combinations of the couplings. 
This is an important result, because it means that a relatively 
low cut -- at about half of the mass of the resonance or even below -- 
is enough to have a good statistical significance, 
and a clear signal from the measurement of the charge asymmetry.

We now fix $m_{t\bar{t}}^{\rm{min}}/m_{G}=0.5$ and 
$m_{t\bar{t}}^{\rm{min}}/m_{G}=0.8$, and we study how the central
asymmetry and its statistical significance vary as a function of
the vector and the axial-vector couplings, 
for a given value of the resonance mass.  
These choices, for which we have found the best 
statistical significances, are of course arbitrary and are 
not necessarily the best for all the values of the vector
and axial-vector couplings. For illustrative purposes are, 
however, good representatives. 
We have chosen $m_{G}=1.5, 2$ and $3$~TeV. 
The results are presented in Figs. \ref{fig:LHC_KK_2d_10TeV} and 
\ref{fig:LHC_KK_2d_14TeV} in the $(g_V,g_A)$ plane for 
$\sqrt{s}=10$~TeV and $14$~TeV, respectively. 
It is possible to see that the pattern of the size of the asymmetry 
is quite similar independently of the value of the resonance mass;
it depends mostly on the ratio $m_{t\bar{t}}^{\rm{min}}/m_{G}$.
A sizable asymmetry is found whatever 
the value of the resonance mass is. The statistical significance, 
as expected, decreases with the increasing of the resonance mass.

\section{Conclusions}

We have analyzed the charge asymmetry in a top-antitop quark pair production 
through the exchange of color-octet heavy boson with flavor independent
coupling to quarks. We have considered the experimental setups of 
Tevatron and LHC, studying different observables and we have found 
that a sizable asymmetry can be found in both.

At Tevatron, the forward-backward asymmetry and 
the pair asymmetry, together with the total 
cross section exclude complementary corners of the parameter space. 
At the LHC, the central charge asymmetry is an
observable that is very sensitive to new physics. 
We have studied the statistical significance of the 
measurement of such an asymmetry, and we have found that  
it is possible to tune the selection cuts in order to find a 
sensible significance.
The maximum of the statistical significance for the measurement 
of the asymmetry as predicted by QCD is obtained without introducing 
any cut in the invariant mass of the top-antitop quark pair, 
even if the asymmetry is smaller in this case.

When a heavy resonance is considered, one or two maxima in the 
significance spectrum are found, depending on the size of the couplings. 
The position of the peaks depends on the ratio 
$m_{t\bar{t}}^{\mathrm{min}}/m_{G}$ and not on the resonance mass. 
One of the peaks can be located at an energy scale as low as one 
half of the resonance mass, or even below. 
Data samples of top and antitop quarks that are not too energetic 
can then be used to detect or exclude the existence of this 
kind of resonances.

\section*{Acknowledgements}

We thank S.~Cabrera, C. Carone, J.H. K\"uhn, M. Sehr, 
and M.~Vos for very useful discussions. 
The work of P.F. is supported by an I3P Fellowship from 
Consejo Superior de Investigaciones Cient\'{\i}ficas (CSIC). 
Work partially supported by Ministerio de Ciencia e Innovaci\'on 
under Grants  No. FPA2004-00996 and CPAN (CSD2007-00042), and 
European Commission MRTN FLAVIAnet under Contract No.  
MRTN-CT-2006-035482.


\appendix

\section{Born cross-section}
\label{ap:born}

The Born cross-section for $q\bar{q}$ fusion in the presence of 
a color-octet vector resonance reads
\bea 
\frac{d\sigma^{q\bar{q}\rightarrow t \bar{t}}}{d\cos \hat{\theta}} &=& 
\alpha_s^2 \: \frac{T_F C_F}{N_C} \:  
\frac{\pi \beta}{2 \hat{s}}
\Bigg( 1+c^2+4m^2 + \frac{2 \hat{s} (\hat{s}-m_G^2)}
{(\hat{s}-m_G^2)^2+m_G^2 \Gamma_G^2} 
\left[ g_V^q \, g_V^t \, (1+c^2+4m^2) + 2 \, g_A^q \, g_A^t \, c  \right] 
\nn \\ &+&
\frac{\hat{s}^2} {(\hat{s}-m_G^2)^2+m_G^2 \Gamma_G^2} 
\bigg[ \left( (g_V^q)^2+(g_A^q)^2 \right)
\bigg( (g_V^t)^2 (1+c^2+4m^2) \nn \\ &+&  (g_A^t)^2 (1+c^2-4m^2) \bigg) 
+ 8 \, g_V^q \, g_A^q \, g_V^t \, g_A^t \, c \, \bigg]
\Bigg)~,
\label{eq:bornqq}
\eea
where $\hat{\theta}$ is the polar angle of the top quark with respect 
to the incoming quark in the center of mass rest frame, 
$\hat{s}$ is the squared partonic invariant mass,
$T_F=1/2$, $N_C=3$ and $C_F=4/3$ are the color factors,
$\beta = \sqrt{1-4m^2}$ is the velocity of the top quark, 
with $m=m_t/\sqrt{\hat{s}}$, and $c = \beta \cos \hat{\theta}$.
The parameters $g_V^q (g_V^t)$, $g_A^q(g_A^t)$ represent the vector
and vector-axial couplings among the excited gluons and the light quarks 
(top quarks). 

There are two terms in \Eq{eq:bornqq} that are odd in the polar 
angle and therefore there are two contributions to the charge asymmetry.
The first one arises from the interference of the SM amplitude with 
the resonance amplitude, and the second one 
from the squared resonance amplitude. 
The former depends on the axial-vector couplings only, while 
the latter is proportional to both the vector and the axial-vector
couplings. For large values of the resonance mass, the second term is 
suppressed, and the charge asymmetry will depend mostly on the 
value of the axial-vector couplings, and residually on the vector 
couplings. The decay width is given by:
\bea
\Gamma_G &\equiv& \sum_q \Gamma (G \to q\overline{q}) 
\approx \frac{\alpha_{s}\, m_G \, T_F}{3} 
\Bigg[\sum_q \left( (g_V^q)^2+(g_A^q)^2 \right) \nn \\
&& +\sqrt{1-\frac{4m_t^{2}}{m_G^{2}}}
\left( (g_V^t)^2 \left(1+\frac{2m_t^{2}}{m_G^{2}}\right) 
+ (g_A^t)^2 \left(1-\frac{4m_t^{2}}{m_G^{2}}\right) 
\right) \Bigg]~.
\eea
We assume that the Born gluon-gluon fusion cross-section 
is the same as in the SM:
\beq
\frac{d\sigma^{gg\rightarrow t \bar{t}}}{d\cos \hat{\theta}} = 
\alpha_s^2 \: \frac{\pi \beta}{2 \hat{s}}  
\left(\frac{1}{N_C(1-c^2)}-\frac{T_F}{2C_F}\right)
\left(1 + c^2 +8 m^2-\frac{32 m^4}{1-c^2}\right)~.
\eeq


\end{document}